\documentstyle[11pt]{article}
 
\def\pa{\partial}

\def\g{\gamma} 
\def\a{\alpha}
\def\b{\beta}
\def\d{\delta} 
\def\e{\epsilon}

\def\l{\lambda} 
\def\m{\mu}
\def\n{\nu}

\def\s{\sigma} 
\def\t{\tau}

\def\mn{{\mu\nu}}
\def\be{\begin{equation}}
\def\ee{\end{equation}}

\begin{document} \thispagestyle{empty} \begin{flushright}
\framebox{\small BRX-TH~533}\\
\end{flushright}

\vspace{.8cm} \setcounter{footnote}{0}

 \begin{center} {\Large{\bf A Note on Stress-Tensors, Conservation and
 Equations of Motion}}\\[8mm]

S. Deser\\
Department of Physics, Brandeis University\\ Waltham, MA 02454,
USA

{\small (\today)}\\[1cm]
\end{center}

\begin{abstract}
Some unusual relations between stress tensors, conservation and
equations of motion are briefly reviewed.
\end{abstract}

Asher Peres is a master of the subtle and unexpected, particularly
in general relativity, his first love.  I therefore present him
with this short collection of (slightly) unlikely relations, from
both sides of the Einstein equations, involving stress tensors,
conservation and equations of motion.

Consider first point systems, whose stress tensors (and all other
properties) are entirely localized along a world line; their
generic form is then
\be
T^\mn (x) = \int ds \left[ a^\m (s) \frac{dx^\n}{ds} + N^\mn (s)
\right] \d^{(m)} (x - x(s)) \; , \;\; N^\mn \frac{dx_\n}{ds} = 0
\; .
 \ee
 This is truly a minimal definition in terms of the {\it a priori}
 undefined vector and
 tensor functions $(a^\m , N^\mn )$ on the world line in dimension
 $m$. The ``surprise", established in relatively recent times \cite{001}, is
that once we demand conservation, $\pa_\m T^\mn = 0$, this
requirement {\it alone} fixes the stress tensor to be that of a
free point particle:
\be
 a^\m = dx^\m /ds \; , \;\; da^\m /ds = 0 \; , \;\; N^\mn (s) = 0
 \; .
 \ee
That is, conservation alone dictates the $T^\mn$ to be just that
of a freely moving point particle.  The existence of an action can
then be inferred uniquely from these ``constraints", to which it
is entirely equivalent.  [As an aside, we mention that demanding
conservation of a similarly localized vector current $j^\m (x) =
\int ds \, v^\m (s) \d^{(m)} (x-x(s))$ yields, as it should, the
weaker result of continuity of the world line described by $v^\m$,
that is, particle conservation.] These results can be extended to
a curved background, by demanding covariant conservation there.

The next surprise is that the same reasoning extends to strings
\cite{002,003}: given a generic $T^\mn$ localized on a
2-dimensional world sheet,
\be
T^\mn(x) = \int d\tau^1 d\tau^2 \tau^\mn (\tau ) \d^{(m)}
(x-x(\tau^a )) \; ,
 \ee
conservation forces $\t^\mn$ to be that of a string,
\be
\tau^\mn (\tau^\m ) = \sqrt{g} \; g^{ab} \; \frac{\pa x^\m}{\pa
\tau^a} \; \frac{\pa x^\n}{\pa \tau^b} \; ,
 \ee
 where $g_{ab}$ is the intrinsic world-sheet metric.
Again conservation has fully determined both content and dynamics
of the system.  The converse surprise \cite{002} is that higher
dimensional, but still localized, higher membranes do {\it not}
share this property of being fully determined by $T^\mn$
conservation.

Can one extend these considerations to fields?  With localization
lost, much more information is required.  For example, if we
specify the stress tensor for a scalar field, say,
\be
 T_\mn = \phi_\m \phi_\n - \textstyle{\frac{1}{2}} \:
 \eta_\mn \; \phi_\a\phi^\a \; , \;\;\;\; \phi_\m \equiv \pa_\m \phi
 \ee
then $\pa_\m T^\mn = \phi_\n \Box\phi$ clearly requires $\Box \phi
= 0$ (which includes $\phi_\n = 0$).  On the other hand, given the
form of $T_\mn$ in (5), we know it is uniquely equivalent to the
metric variation of the normal scalar action (evaluated at flat
space, for simplicity).  From this it follows that $T^\mn$ is
conserved on- and only on-shell, since the action
 \be
 I = -\textstyle{\frac{1}{2}} \: \int (dx) \sqrt{-g} \: g^\mn
 \phi_\m\phi_\n \; ,
\ee
 is diffeomorphism invariant under arbitrary gauge functions
$\xi_\m (x)$; hence
\be
 \d I = 0 = \int (\xi_\m(x) \: D_\n \: T^\mn + \int (\d_\xi \phi )
 \frac{\d L}{\d\phi}
\ee
 and the last term is the field variation contribution.  Effectively then,
 given a form for $T^\mn$, we know it is only of interest if
 derivable from an action, in which case we are in the usual
 circumstances of field theory.  [If we had instead chosen a ``bad"
 $T_\mn$, say $T_\mn =  \phi_\m \phi_\n$, then $\pa_\m T^\m_\n =
 \phi_\n \Box \phi + \frac{1}{2} \pa_\n \phi^2_\a$, whose
 vanishing
leaves no interesting dynamics.]

There is an additional sublety, depending on whether the action is
in first or second order form, as we illustrate (for a change)
through  the vector field.  If we assume the usual form of
$T_\mn$,
\be
T_\mn = F_{\m\a} F_\n~\!^\a - \textstyle{\frac{1}{4}} \: g_\mn \:
F^2_{\a\b} \; , \ee
 and that $F_\mn$ is a curl, then conservation of $T_\mn$ of
 course reduces to the requirement
 \be
 F_{\m\a} \pa_\b F^{\b\a} = 0 \; .
 \ee
Since (7) is also the metric variation of the (second-order)
Maxwell action,
\be
I_M = - \textstyle{\frac{1}{4}} \int \sqrt{g} \:g^{\m\a}g^{\n\b}
F_\mn F_{\a\b} \; , \ee
 this simply parallels the scalar discussion.  However, we also know
 that there is a first order form of $I_M$, with $(A_\m , \:
 F^\mn )$ independent,
 \be 
 I_M [ A,F] = - \textstyle{\frac{1}{2}} \int (dx)  [F^\mn (\pa_\m A_\n
 - \pa_\n A_\m ) -\textstyle{\frac{1}{2}} F^{\m\a} F^{\n\b} g_\mn
 g_{\a\n} \; \sqrt{-g}\; ] \; .
 \ee
In this form, both $\pa_\m F^\mn = 0$ {\it and} $F_\mn = (\pa_\m
A_\n - \pa_\n A_\m )$ are on a par as field equations, and the
symplectic $F\pa A$ term does not contribute to the metric
variation ($F^\mn$ is here a contravariant density).  The Maxwell
tensor form (7) remains, however, but its conservation now
requires that both these field equations hold, since $F^\mn$ in
(7) is now just an algebraic variable.

Returning to the condition (9) in second order form, obtaining the
Maxwell equation from it requires proving that it implies $\pa_\b
F^{\b\m} = 0$, {\it i.e.}, that the determinant of its
coefficient, det$|F_{\a\b}|$, vanish.  Now this determinant is of
the generic form $\a (E^2 - B^2)^2 + \b (E\cdot B)^2$, and
vanishes for plane waves, $E\cdot B = 0, \; |E|=|B|$.  So
generically, stress tensor conservation implies field equations
(which plane waves also obey) for the Maxwell system too.

More generally, one might draw the usual conclusion that for any
matter system describable by a diffeo-invariant action, hence
endowed with a well-defined $T^\mn$ (up to the usual identically
conserved superpotential terms), its conservation is generically
equivalent to satisfaction of the corresponding field equations.
But, there is yet another twist on the stress tensor--action
connection.  Two inequivalent actions can share the same $T^\mn$ !
One example is any term, such as the Chern--Simons (CS) invariant,
whose action is topological {\it i.e.}, metric-independent.  Hence
it does {\it not} contribute to $T^\mn$ at all, but it had better
-- and does -- preserve conservation; how?  The simplest example
is $D$=3 vector gauge theory with CS term; here is the abelian
version:
\be
 I_{TME} = - \frac{1}{4} \int d^3x \, F^2_\mn + m/2 \, \int \,
 d^3x \, \e^{\mn\a} A_\m F_{\n\a}\; .
 \ee
Conservation of the normal Maxwell tensor (8) is ``saved" by an
identity.  The field equation $\pa_\m F^\mn + m\e^{\n\a\b}
F_{\a\b} = 0$ implies that
 \be
 F_{\n\a} \pa_\b F^{\b\a} = - F_{\n\a} \, m \, \e^{\a\l\s}F_{\l\s}
 \equiv 0 \; .
  \ee
  Other such ``topological" terms include, in $D$=4,
  the truncated Born--Infeld action $\int (d^4x)
  \sqrt{\det \, F_\mn}$~, which
  is also clearly metric-independent.  In this sense, stress
  tensor conservation need not fully determine the action it
  represents -- just the opposite of our initial localized
  examples.

Our final case is the gravitational field itself, which of course
has no well-defined stress tensor.  To what extent, then, can we
construct parallel results to those valid for matter?  Clearly,
{\it any} geometric scalar density Lagrangian
\be
I=\int (dx) \sqrt{-g} \: L (R, \: R^2, \: (DR) \ldots ) \; ,
 \ee
 has identically diffeo-invariant action, implying the (generalized)
 Bianchi identity for its metric variation, $\d I/\d g_\mn$.
 But we are not interested in identically conserved
 quantities.  This is how the only other tensor available, that of
 Bel--Robinson, entered Einstein gravity. We recall in $D$=4,
 \be
 T_{\a\b\g\d} = R_{\a\rho\g\s}\; R_\b~\!^\rho_\d\!~^\s + \;
 ^*\!R_{\a\rho\g\s} \; ^*\!R_\b\!~^\rho_\d~\!^\s
 \ee
is (covariantly) conserved by virtue of Ricci flatness; indeed
 that was its original appeal.  In its cleanest form, with the
 curvature in (14) replaced by Weyl tensors, conservation
 clearly depends only on derivatives (say curls) of the Ricci
 tensor, rather than on $R_\mn$ or $G_\mn$ algebraically.  Then
 [even apart from the plane-wave determinant ambiguities akin to
 the Maxwell case from $\sim R(DR) = 0$] this means that weaker
 field equations, namely those where $R_{\m[\a ;\b ]} = 0$ also conserve
 the same $T_{\mn\a\b}$.  This is amusing since these latter field
 equations are not derivable as a Lagrangian system, though they admit
 Einstein spaces.

 There is however, something a bit strange here compared to lower
 spins.  Since $BR$ is a 4-index quantity unrelated to any gauge
 symmetry, it is really more like an ``invariantization" of the
 non-invariant stress tensor by applying more derivatives to it.
 Indeed, we can draw the following parallel with Maxwell theory.
 One may define a 4-index tensor there \cite{004} which is
 essentially of the form
 \be
 \t_{\mn\a\b} \sim \pa_\a F_{\m\l}\pa_\b F_\n^{~\!\l} + \ldots
 \ee
 and consequently is also conserved on Maxwell shell, as is its
 more transparent trace
 \be
\t_\mn \sim \pa_\a \: F_{\m\l} \: \pa^\a F_\n^{~\!\l} -
 \textstyle{\frac{1}{4}} \: \eta_\mn (\pa_\a F_{\b\g})^2
 \ee
 But in fact this quantity (in flat space) is just the wave operator
 acting on the Maxwell  tensor (on-shell, which
 is all that counts, $\Box F_\mn =0$),
 \be
 \t_\mn = \textstyle{\frac{1}{2}} \: \Box \; T^{max}_\mn \; ,
 \ee
 So such tensors, while rather superfluous for lower $(s \leq 1)$ spins, are really
 needed for higher ones.  Indeed, all higher than $s$=1 gauge fields,
 already at the tree level, have non-gauge invariant
 stress-tensors \cite{006} and would require their own BR extensions, with
 more and more derivatives to emulate BR.

 In summary, we have reviewed how stress tensors are surprisingly
 powerful for point and string -- but not higher brane --
 systems: localized but otherwise
 arbitrary, their conservation suffices to specify all the
 dynamics.  For lower spin fields, their actions specify the
 stress tensors but {\it not} the converse, as
 illustrated by the CS example:  It is ambiguous to state
  that conservation occurs
on -- and only on -- shell, since there can be more than one
``shell". For gravity theories, there is no independent stress
tensor but only a ``covariantized", BR, version necessarily
involving higher derivatives, hence requiring somewhat weaker
conditions than Ricci-flatness for (covariant) conservation.
Similar considerations hold also in attempting to endow other high
spin gauge fields with gauge invariant stress tensors. These are
however rather different conditions than for the -- fully
covariant -- gravity case.  For example, free massless spin $>$1
fields also have non-invariant stress tensors, but their
conservation is as directly linked to the (invariant) field
equations as for lower spins.

The moral is that, while the standard lore is always essentially
correct, there is always some fine print to take into account.

This work was supported by the National Science Foundation under
grant PHY99-73935.


\begin{thebibliography}{999}
\bibitem{001}
A.\ Tulczjiew, Acta Phys.\ Pol.\ {\bf 18} (1939)  393.
\bibitem{002}
C.\ Aragone and S.\ Deser, Nucl.\ Phys.\ {\bf B92} (1975) 327.
\bibitem{003}
M.\ G\"{u}rses and F.\ G\"{u}rsey. Phys.\ Rev.\ {\bf D11} (1975)
967.
\bibitem{004}
M.\ Chevreton, Nuovo Cim.\ {\bf 34} (1964) 90.
\bibitem{005}
S.\ Deser, Class.\ Quant.\ Grav.\ {\bf 20} (2003) L213.
\bibitem{006}
S.\ Deser and J.\ McCarthy,  Class.\ Quant.\ Grav.\ {\bf 7} (1990)
L119; S.\ Deser  and A.\ Waldron, in preparation.
\end{thebibliography}
\end{document}